\documentclass[aps,preprint,showpacs]{revtex4}

\usepackage{amsmath}
\usepackage{graphicx}
\usepackage{dcolumn}
\usepackage{bm}
\usepackage{amssymb}

\makeatletter
\def\@biblabel#1{#1}
\makeatother

\begin{document}

\title{Common dependence on earthquake magnitudes for the trapped particles bursts approaching the earthquake}
\author{Ping Wang, Huanyu Wang, Yuqian Ma, Hong Lu, Xiangcheng Meng, Jilong Zhang, Hui Wang, Feng Shi, Yanbing Xu, Xinqiao Li, Xiaoxia Yu, Xiaoyun Zhao, Feng Wu, Zhenghua An, Wenqi Jiang, Hanyi Liu}
\affiliation{Institute of High Energy Physics,Chinese Academy of
Science, 100049 Beijing, China }
\email{pwang@ihep.ac.cn}

\begin{abstract}
\begin{center}

  Trapped particles bursts have long been observed to be frequently occurred several hours before earthquakes, especially for strong earthquakes, from several space experiments during past decades. However, the validity of earthquake origin of particles bursts events is still unsolved. In this paper, we firstly reported the frequency distribution and time evolution of particles bursts for various magnitudes, which within different time windows centered around earthquakes. The results showed nearly the same systematic dependence of particle bursts frequency on earthquake magnitude and characteristic time evolution behavior of average number of particles bursts for various magnitudes. These findings should strengthen the validity of earthquake origin of particles bursts and further understanding of particles bursts as possible precursor of earthquake.

\end{center}
\smallskip
\end{abstract}


\maketitle


A variety of premonitory phenomena associated with earthquakes exhibit anomalous effects which believed to be correlated with seismic activity, such as mechanical deformation, geochemical and hydrological precursors, and electromagnetic precursors. Recently, radiation belt energetic particle fluxes showed promising sign of precursor of strong seismic activity from past several space experiments\cite{Galper-1, Aleksandrin, Sgrigna, Pulinets,Fidani}. These particle fluxes are characterized by an anomalous short-term and sharp increase of high-energy particle counting rates which are referred as particle bursts (PBs).

PBs arises when subjected to electromagnetic disturbances in space environment. For inner radiation belt, pitch angle diffusion plays dominant role compared with other processes\cite{Abel}, where trapped particles are scattered into loss cone and result in PBs events and particle precipitation. Practically, the occurrences of PBs events are frequently influenced by many natural phenomena, such as thunderstorm or geomagnetic storm\cite{Blake, Reeves, Horne, Inan}, which could result in that earthquake origin of PBs events are difficult to distinguish from those of non-seismic sources. Furthermore, many earthquakes indeed do not accompanied by PBs events within time intervals between them, namely time windows, from several hours to several days. So the validity of earthquake origin of PBs remains largely unsolved.

In this research, we investigated the frequency of PBs occurrence centered around earthquakes within different time windows for various magnitudes, and found essentially nearly the same systematic dependence of PBs frequency on earthquake magnitude and its characteristic time evolution behavior. These findings indicated that PBs events approaching earthquakes are positively correlated with earthquake magnitude and its average number is uniformly decreased with time from the beginning of earthquake events. Our results more directly related the PBs events with earthquake and should strengthen the validity of earthquake origin of PBs events.


To resolve the issue of the validity of earthquake origin of the PBs events, we searched the Data set collected from the DEMETER experiment. The DEMETER (Detection of Electro-Magnetic Emissions Transmitted fro Earthquake Regions) satellite is the first of its kind in the micro satellite series, which aims at the research for the space electromagnetic and high energy particle precursors simultaneously associated with strong seismic events\cite{Sauvaud}. For trapped radiation belt observation, the DEMETER worked just within loss cone or nearby for its orbit and location of particle detector - IDP (The Instrument Detecteur de Particules)\cite{Sauvaud}.

In the present paper we used the data set ranged in time from 2005 to 2010 including electron energy $E=0.97-2.3$ MeV channel (possible protons are mixed with electrons in this energy range) under survey mode of the DEMETER experiment, together with seismic data were supplied by the IPGP (Institut De Physique Du Globe De Paris). In particularly, We focused on the range McIIwain L-parameter $1.3\leq L\leq 1.4$ both with seismic events and trapped particles, respectively, which is near the lower boundary of inner radiation belt. For much lower L values ($L<1.3$), collisions of electrons with atmospheric atoms dominate the loss of trapped electrons\cite{Walt} while rarely counting rates were recorded at much higher L value ($L>1.4$) with energy in previous range during this experiment.

In this research, the PBs was defined as current counting rates which exceeded 4 standard deviations from average value of the background flux ,with high counting rates region (including the South Atlantic Anomaly (SAA) region) $(lat:-90^{\circ}\sim 0^{\circ}, lon:-100^{\circ}\sim 45^{\circ})$ were excluded from consideration to decrease the background flux influence. As in survey mode all orbit related parameters were provided every 28 seconds, so counting rates were averaged over 28 seconds and each one PBs event at least experienced duration of 28 seconds. For one PBs event lasting over 28 seconds and still staying in the region $1.3\leq L\leq 1.4$, it will be regarded as two or more PBs events depending on PBs time duration.

Considering temporal correlation between PBs events and seismic events, an example of PBs events observation were carried out within a time window of $\pm1$ day for magnitude greater than or equal to 5.0 during Jan of 2005 [see Fig. 1]. In this case each earthquake was corresponded to one or more PBs events, but there were still many PBs events that did not correspond to anyone of earthquakes within selected time window. It should be noted that the possibility of one PBs event may be regarded as two at the same time if one PBs event occurs just within the overlapping of two time windows which belonging to two successive seismic events. However, this case is rarely happened in our research and should not introduce evident systematic errors.

\begin{figure}[h]
 \begin{center}
 \includegraphics[scale=0.7]{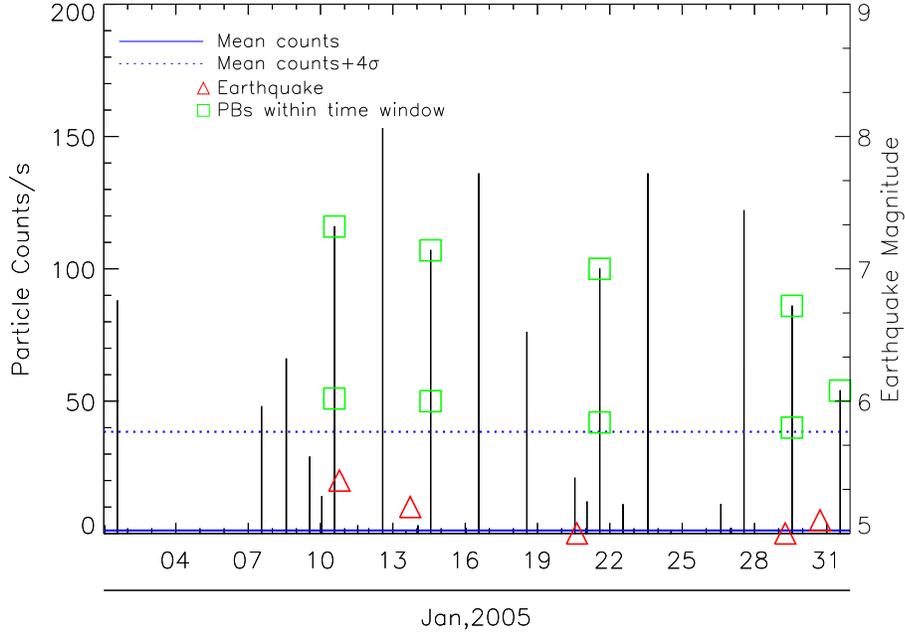}
 \caption[]{Time Correlation between PBs events and seismic events. The solid blue line is the average counts
  of particles during Jan, 2005.
  The dotted blue line is the sum of the average counts and $4\sigma$, where $\sigma$ is the standard
  deviation. The green squares are PBs events within a time window $\pm 1$ day. Seismic events are represented by red triangles.}
 \label{}
 \end{center}
 \end{figure}

In order to investigate earthquake origin of the PBs events more comprehensively, earthquakes are classified according to magnitude M from $M\geq 5.0$ to $M\geq 6.2$, in which intervals between 5.0 and 6.2 are 0.3, 0.4 and 0.5, respectively. (Note that the magnitude intervals are not of equal width but were chosen to have a significant number of events in each interval.) The number N is the total events with magnitude greater than or  equal to M, which is logarithmically decreased as magnitude M is linearly increased. As magnitude is increased from $M=5.0, 5.3, 5.7$ to $6.2$, the numbers of seismic events(from IPGP) are decreased from $N=385, 166, 79$ to $27$ correspondingly, with both events distributed between $1.3\leq L\leq 1.4$ during selected period. More stronger earthquakes greater than 6.2 were not classified in present research in order to decrease the influence caused by fluctuation for rapidly decreased numbers of seismic events and increase the reliability of analysis.

To further determine whether the PBs events are originated from earthquakes, we studied the variation of the occurrence of PBs events, especially its frequency distribution within different time windows for various magnitudes. The frequency distribution $P(k)$, where $k$ is the frequency of PBs events, exhibited distinctive behavior according to magnitudes within time window $\pm0.5$ day [see Fig. 2(a)] and evolved to completely different one within time window $\pm4$ days [see Fig. 2(b)]. The results [see Fig. 2(a)] obviously showed that many earthquakes, regardless of magnitude, are not accompanied by PBs events ($k=0$) at this time interval. However, the data for different magnitudes showed essentially nearly the same systematic dependence of PBs frequency on earthquake magnitude within the time window $\pm0.5$ day [see Fig. 2(a)].

The absolute value of PBs probability for different frequencies of PBs occurrence is not concerned here. The essential result is that, in all cases, the larger magnitude has the relative higher level of PBs occurrence ($1-P(0)$). As earthquake magnitude was increased from 5.0 to 6.2, the probability of the PBs occurrence ($1-P(0)$) was monotone increased from $36\%$ to $56\%$. More detailed information were showed for different nonzero k ,which also exhibited similar systematic dependence just liked the case $k=0$. In some data sets, as $k$ increased, separation between different magnitudes was no longer clear or even deviated from systematic dependence for the cases of the largest $k$ value or for the highest earthquake magnitude because of fluctuation introduced by small sample sizes.

The average number of PBs events in various time windows was obtained for per earthquake and per day(time window). The results exhibited similar decay form [see Fig. 2(c)] for different magnitudes with time windows extended from 0.5 to 4 days. The separation is clear for different magnitudes and the decay form is specially evident for magnitude greater than $5.0$.

 \begin{figure}[h]
 \begin{center}
 \includegraphics[scale=0.6]{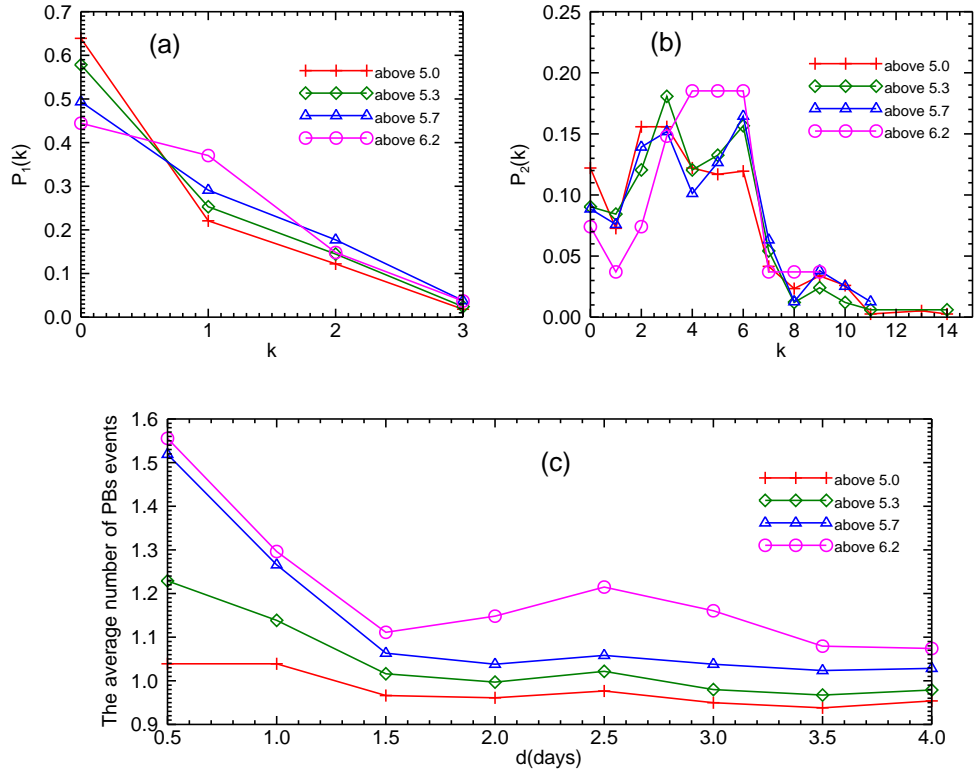}
 \caption[]{The probability distribution of PBs frequency and the time decay behavior of the average number of PBs events within different time windows for various earthquake magnitudes. (a) The probability distribution of PBs frequency $P_1(k)$ within a time window $\pm 0.5$ day for various earthquake magnitudes, where k is the number of PBs events. (b) The probability distribution of PBs frequency $P_2(k)$ within a time window $\pm 4$ days for various earthquake magnitudes. (c) The time decay behavior of the average number of PBs events within different time windows for various earthquake magnitudes. Symbols in (b) and (c) are all according to the legend in (a).}
 \label{}
 \end{center}
 \end{figure}

The data sets of particles selected are very different in space environment and geographical distribution except excluding of SAA region. Nevertheless, all data sets exhibit the same dependence on earthquake magnitude both for the probability of PBs occurrence and the average number of PBs events for per earthquake and per day(time window).

According to the analysis described above, the uniformity of the same systematic dependence of PBs frequency on earthquake magnitude needs a universal interpretation. This requirement is strengthened by the systematic dependence of the average number of PBs events on earthquake magnitude and time window. We suggest that PBs events, at least within the time window $\pm0.5$ day, are originated from seismic events. This conclusion is illustrated by the observation of positive correlation between frequency of PBs occurrence and earthquake magnitude. Furthermore, the average number of PBs events also showed positive correlation with earthquake magnitude and time window. Our findings indicated that the complexity of strong seismic events may be characterized by the variation of the PBs frequency and average number decay behavior of PBs events.

The underlying mechanisms responsible for earthquake origin of the PBs events may be better understood by considering the example of radiation belt electron precipitation by ground-based transmitters. Electromagnetic wave generated by ground-based transmitters can penetrate through ionosphere and propagate with such as whistler or other modes along geomagnetic field. When they encounter energetic electrons in radiation belt, wave-particle interaction may be happen and pitch angle diffusion will play dominating role during this process, especially for inner radiation belt\cite{Abel}. Then, sharp and short-term increases of particle counting rates, as PBs events, occur after electrons are scattered into loss cone and precipitate into higher atmosphere eventually. Analogously, similar process may be happen if earthquake really produce electromagnetic wave directly or indirectly. They can arrive firstly at the surface of the Earth and then propagate up to the space where they encounter energetic electrons. The possibility of above assumption has been discussed over past decades\cite{Molchanov-1, Molchanov-2}. Since trapped electrons in radiation belt drift along nearly the same L shell as epicenter before they precipitate into higher atmosphere, the instrument on board a satellite is able to collect earthquake information globally during one half orbit.

Previous studies\cite{Aleksandrin, Sgrigna}, which from a variety of consistent observation and based on more higher energy range of electrons or protons, showed that peaks of PBs events were more frequently appeared a few hours before earthquakes with magnitude greater than $4.0$ while the most relevant L shell range is about $0.07$. This promising feature of PBs events makes it possible as reliable precursor of earthquake. Compared with previous works, which focused on the space and time correlations between PBs events and seismic events, we emphasized both on the correlation between the frequency of PBs events and magnitude as well as on the time evolution behavior of average number of PBs events. Our results directly related the PBs events with earthquake magnitude and clearly showed the characteristic behavior of PBs events approaching earthquake. Considering this situation, the validity of earthquake origin of PBs events may be better convinced.

We note that although our findings about the common dependence of PBs events on earthquake magnitudes and time windows are presented only for electrons energy in the range $E=0.97-2.3$ MeV, we have not shown that these features are unique to this energy channel. In fact, the electrons in the lower energy channel $E=0.52-0.97$ MeV behave similarly in most cases. However, the electrons in the lowest energy channel $E=0.09-0.52$ MeV do not exhibit similar behavior. One possible reason may be the important role played by ground-based transmitters in the process of inner radiation belt electrons precipitation\cite{Abel, Inan-1, Li}. Trapped electrons, which belong to this lowest energy channel, are greatly influenced by cyclotron resonance with VLF wave produced by ground-based transmitters.

It should be noted that this study has examined only for area $1.3\leq L\leq 1.4$ and the data sets from the DEMETER experiment, more choices of L range and extensive data sets are needed to verify the common dependence of PBs events on earthquake magnitudes and time windows. Furthermore, more choices of L range are helpful to determine the effective L interval between PBs events and seismic events which affect the result of the dependence conversely. Presently, contributions to PBs events from other natural phenomena such as thunderstorm and geomagnetic storm are not excluded, so the results can not be used to determine the absolute value of PBs events associated with earthquake. Moreover, the influence from different hypocentral depth\cite{Pustovetov} and oceanic crust earthquakes\cite{Oike, Galper-2, Parrot} are not took into consideration. Despite its preliminary character, the validity of earthquake origin of PBs events are still clearly showed and systematic deviation are not introduced by those of non-seismic sources from uniform distribution of PBs events and similar time evolution behavior of average number of PBs events.

The possibility of PBs events as reliable precursor of earthquake was not addressed in these studies. In fact, a natural extension of the present works would be to investigate whether the time decay behavior of average number of PBs events is still valid according to different time windows for various magnitudes just before earthquake. Future studies should also include a parallel analysis in possibility of providing an alternative method for quantifying magnitude of earthquake by PBs characteristic time evolution behavior. Compared to the procedure that 'magnitude scale' is defined, which firstly established by Richter in 1935, they share the similar decay form for various magnitudes. But the most essential difference between them may be that our result is valid for average number of PBs events within selected time windows, while 'Richter scale' is set up by measuring the largest amplitude of displacement from the epicenter.

We have shown that the uniformity of the same systematic dependence of PBs frequency on earthquake magnitudes, as well as average number of PBs events on time windows, which both clearly demonstrate the validity of the earthquake origin of the PBs events. Clearly, much more work will need to be done to fully demonstrate the validity in future. Further understanding of PBs variation before earthquake is necessary for the future development and prosperity of reliable precursor of earthquake.

We thank M. Parrot for valuable comments. We thank the DEMETER and the CDPP (Centre de Donn谷es de Physique des Plasmas) for providing of data.

\end{document}